\documentstyle[12pt,aasms4]{article}

\def\degree{\ifmmode {^\circ}\else {$^\circ$}\fi}
\def\rstar{\ifmmode {\, R_{\star}}\else $R_{\star}$\fi}
\def\msol{\ifmmode {\, M_{\odot}}\else $M_{\odot}$\fi}
\def\rsol{\ifmmode {\, R_{\odot}}\else $R_{\odot}$\fi}
\def\lsol{\ifmmode {\, L_{\odot}}\else $L_{\odot}$\fi}
\def\msolyr{\ifmmode {\,M_{\odot}\,{\rm yr}^{-1}}\else $M_{\odot}\,{\rm yr}^{-
1}$\fi}
\def\mdot{\ifmmode {\,\dot{M}}\else $\dot{M}$\fi}
\def\mdotyr{\ifmmode {\,\dot{M}\,yr^{-1}}\else $\dot{M}\,yr^{-1}$\fi}

\begin{document}

\title{The MACHO Project LMC Variable Star Inventory: IV.\\ New R Coronae Borealis 
Stars} 
\author {C. Alcock$^{1,2}$, R.A. Allsman$^1$, D.R. Alves$^{1,3}$,
T.S. Axelrod$^1$, A. Becker$^{2,4}$, D.P. Bennett$^{1,2}$, G.C. Clayton$^{5}$, 
K.H. Cook$^{1,2}$, 
K.C. Freeman$^6$, K. Griest$^{2,7}$, J.A. Guern$^{2,7}$, D. Kilkenny$^8$, M.J. 
Lehner$^{2,7}$, S.L. Marshall$^{2,9}$, D. Minniti$^1$,
B.A. Peterson$^6$, M.R. Pratt$^{2,9}$, P.J. Quinn$^6$, A.W. Rodgers$^6$, C.W. 
Stubbs$^{2,4,9}$, W. Sutherland$^{10}$, and D.L. Welch$^{11}$}
\affil { (The MACHO Collaboration) }

\altaffiltext{1}{Lawrence Livermore National Laboratory, Livermore, CA 94550\\
E-mail:  alcock, robynallsman, alves, tsa, bennett, kcook, dminniti@llnl.gov} 

\altaffiltext{2}{Center for Particle Astrophysics, University of California, 
Berkeley, CA 94720}

\altaffiltext{3}{Department of Physics, University of California, Davis, CA 95616}

\altaffiltext{4}{Department of Astronomy, University of Washington, 
Seattle, WA 98195 \\
E-mail: stubbs@welkin.astro.washington.edu, becker@astro.washington.edu}
 
\altaffiltext{5}{University of Colorado, CASA, Campus Box 389, Boulder, CO 80309\\ 
E-mail: gclayton@fenway.colorado.edu}
 
\altaffiltext{6}{Mt.  Stromlo and Siding Spring Observatories, Australian National 
University, Weston, ACT 2611, Australia \\ E-mail: kcf, peterson, pjq, 
alex@merlin.anu.edu.au}
 
\altaffiltext{7}{Department of Physics, University of California, 
San Diego, CA 92093 \\
E-mail: griest, jguern, matt@astrophys.ucsd.edu}

\altaffiltext{8} {South African Astronomical Observatory, P.O. Box 9, Observatory 7935, 
South Africa\\
E-mail: dmk@da.saao.ac.za}
 
\altaffiltext{9}{Department of Physics, University of California, 
Santa Barbara, CA 93106 \\
E-mail: stuart, mrp@lensing.physics.ucsb.edu}
  
\altaffiltext{10}{Department of Physics, University of Oxford, 
Oxford OX1 3RH, U.K.\\
E-mail: wjs@oxds02.astro.ox.ac.uk}
 
\altaffiltext{11}{Department of Physics and Astronomy, McMaster University,
Hamilton, ON L8S 4M1, Canada\\
E-mail: welch@physics.mcmaster.ca}

\begin{abstract}
We report the discovery of two new R Coronae Borealis (RCB) stars in the Large 
Magellanic Cloud (LMC) using the MACHO project photometry database.  The 
identification of both stars has been confirmed spectroscopically.  One is a cool RCB star 
($T_{eff}\sim$ 5000 K) characterized by very strong Swan bands of $C_2$ and violet 
bands of CN, and weak or absent Balmer lines, G-band and $^{12}C^{13}C$ bands. The 
second star is an example of a hot RCB star of which only 3 were previously known to 
exist in the 
Galaxy and none in the LMC.  Its spectrum is characterized by several C II lines in 
emission.  Both stars have shown deep declines of $\Delta V \ge 4$ mag in brightness.  
The new stars are significantly fainter at maximum light
than the three previously known LMC RCB stars.  
The amount of reddening toward these stars is somewhat uncertain but both seem to have  
absolute magnitudes, $M_V$, about half a magnitude fainter than the other three stars.
Estimates of $M_{Bol}$ find that the hot RCB star lies in the range of the other three stars 
while the cool RCB star is fainter.  The two cool LMC RCB stars are the faintest at 
$M_{Bol}$.
The discovery of these two new stars brings to five the number of known RCB stars in the 
LMC and demonstrates the utility of the MACHO photometric database for the discovery of 
new RCB stars.

\end{abstract}

\section{Introduction}
The R Coronae Borealis (RCB) stars represent a rare type of hydrogen-deficient 
carbon-rich supergiants which undergo very spectacular declines in visual brightness of 
up to 8 magnitudes at apparently irregular intervals (Clayton 1996).  
A cloud of carbon-rich dust forms along the line of sight to 
the RCB star eclipsing the photosphere, causing a severe drop in its brightness and the 
appearance of a rich emission-line spectrum.  As the dust cloud 
disperses, the star returns to maximum light.  
RCB stars have a wide range of temperatures but they can be divided simply into three 
groups, cool ($\sim$5000 K), warm ($\sim$7000 K) and hot ($\sim$20,000 K).  Typical 
representatives of 
these groups are S Apodis, R Coronae Borealis and V348 Sagittarii, respectively.  Most 
RCB stars fall in the warm category. 
Hot RCB stars are quite rare with only 3 examples known. 
The typical warm RCB spectrum at maximum light looks like an F or G supergiant with a 
few important differences: 
the Balmer lines are very weak or absent; the spectrum contains many lines of neutral 
carbon, and bands of $C_2$ and CN. 
The cool RCB-type spectrum resembles the warm type but with much stronger molecular 
absorption bands.  
The hot RCB stars show similar lightcurve behavior to the cooler stars but their spectra are 
very different (Pollacco \& Hill 1991).  The spectrum of V348 Sgr, the best studied 
hot-type star, shows strong emission lines of C II and He I as well as the Balmer lines, Ne 
I and various forbidden lines (Dahari \& Osterbrock 1984).
Most RCB stars in all three categories show excesses at near-IR and IRAS wavelengths.
 
The RCB Stars are very rare either because they form 
only in unusual circumstances or because they are a brief episode in stellar 
evolution.  
Only about 30 RCB stars are known in the Galaxy, and until now only 3 in the LMC 
despite their high intrinsic luminosities.
Their evolutionary history remains very uncertain.  
Two major evolutionary models have been suggested for the origin of RCB stars, the 
Double Degenerate and the Final Helium Shell Flash conjectures 
(Sch\"{o}nberner 1986; Renzini 1990; Iben, Tutukov, \& Yungelson  1996). Both 
involve expanding white dwarfs to the 
supergiant sizes assumed for RCB stars.  
A third model suggests that RCB stars are binaries in the second common envelope 
phase with a low mass companion orbiting inside the envelope (Whitney, Soker, \& 
Clayton 1991).
Recently, Iben et al. (1996) added the merger of a neutron star and a helium-rich star to the 
list of possible RCB star precursors.

An important input parameter to these models is stellar luminosity. This parameter can only 
be estimated when the distance to a star is known.
However, there is no reliable distance estimate to any Galactic RCB star.
Since they are not ``normal" stars, their distances can only be estimated if they are 
associated with an object at a known distance or through other indirect methods.
Previous estimates of Galactic RCB star luminosities are summarized in Table 1.
In addition, a star in the cluster NGC 6231 was initially identified as an RCB star but 
turned out to 
be a normal reddened star (Bessel et al. 1970; Herbig 1972).
In a similar manner to Doroshenko et al. (1978), Rosenbush (1981, 1982, 1989, 1995), using 
estimates of reddening along sightlines to RCB stars and the structure of the interstellar 
medium,  finds a wide range of absolute magnitudes,  $M_V$ =  -5 to +2.5.
The RCB star,  V482 Cygni was identified with a quadruple star system containing a K5 
III star based on proximity on the sky implying $M_V$ = -2.8 (Gaustad et al. 1988).  
This association was refuted by Rao \& Lambert (1993) who find that V482 Cyg has 
significantly different radial velocities and interstellar columns than the K5 III star.  They 
estimate a larger distance consistent with an $M_V \sim$ -4.6.
Other RCB stars, including RY Sagittarii,  have close companions although none have been 
shown to be physical pairs (Andrews et al. 1967, Feast 1969; Milone 1995).
Estimates of Galactic RCB star luminosities differ by factors of up to $10^3$.

Due to the absence of reliable distance estimates for the Galactic stars, the LMC RCB stars 
play a pivotal role in RCB star research.
Absolute luminosities can be derived from the LMC RCB stars which are at a known 
distance.
Using their 
apparent magnitudes and the known distance of the LMC (m-M = 18.6), an absolute 
magnitude of $M_V\sim-4~to -5$ is derived.  However, this is based on only 3 stars (Feast 
1972).
This result, that RCB stars have supergiant size and luminosity, puts strong constraints on 
the evolutionary models outlined above.

One of the dividends from the search for Massive Compact Halo Objects (MACHO's) 
towards 
the LMC is the discovery of a large number of new variable stars.  Over 40,000 variables 
have been discovered so far (Cook et al. 1995).
RCB candidates have been selected on the basis of their lightcurve behavior and confirmed 
spectroscopically. When only fragmentary lightcurve data are available, RCB stars may be 
confused with 
symbiotic, cataclysmic or semi-regular variables (Lawson \& Cottrell 
1990). 

\section{Known LMC RCB Stars}
Outside the Galaxy, only 3 RCB stars have been discovered to date, W Mensae, HV 5637, 
and HV 12842 
(Rodgers 1970; Payne-Gaposchkin 1971; Feast 1972).  
They are thought to be members of the LMC.
Radial velocities for HV 12842 and W Men are appropriate for LMC membership (Feast 
1972; Pollard, Cottrell, \& Lawson 1994).
The radial velocity of HV 5637 is not known.
HV 12671 was previously identified as an RCB star but is now thought to be a 
carbon-symbiotic star (Allen 1980; Lawson et al. 1990).
The three LMC RCB stars are listed in Table 2.
Photometric coverage has been spotty but declines have been observed for each of the 
stars.  
HV 5637 only has one decline on record and no IR excess (Glass, Lawson, \& Laney 
1994). It may be 
similar to the Galactic RCB star, XX Camelopardalis (Clayton 1996).
The lightcurve behavior of the three stars is summarized in Lawson et al. (1990).  
Long-term B and V photometry was obtained by Lawson et al. for 
W Men and HV 12842.  A few observations of HV 5637 were also obtained.  
The V magnitudes at maximum light ($V_{max}$) are listed in Table 2. 
The lightcurves of W Men and HV 12842 demonstrate small amplitude variations similar to 
those typically 
seen in the Galactic stars.
Spectra of all three stars were obtained by Feast (1972). The spectra of W Men and HV 
12842 show that they belong in the warm RCB group, very similar to R CrB and RY Sgr, 
showing weak $C_2$ bands (Rodgers 1970; Feast 1972).  HV 5637 is a cool RCB star 
with a spectrum similar to S 
Aps having very strong bands of $C_2$.
The B-V of HV 5637 implies a spectral type of K2 (Glass et al. 1994). For W Men and HV 
12842 the B-V colors indicate mid-F.

Eggen (1970) points out that the U-B colors are quite a bit bluer for W Men (an LMC RCB 
star) than for RY 
Sgr or R CrB (Galactic RCB stars).
Pollard et al. (1994) have measured fine abundances for 
HV 12842 and W Men.  They are similar in composition to the majority of Galactic RCB 
stars except that they are iron deficient (Lambert \& Rao 1994).  This is perhaps not 
surprising since, in general, stars in the 
LMC are iron deficient. 
The RCB stars are characterized by extreme hydrogen deficiency and an overabundance of 
carbon. 
In general, C/H $\geq 10^3$, [C/Fe] $\sim 1$, [X/Fe] $\sim$ solar for most other species 
up to iron peak elements and $^{12}C/^{13}C \geq 100$ (Pollard et al. 1994). 
The high  $^{12}C/^{13}C$ ratio implies the presence of material processed by 
helium-burning.
Lambert \& Rao (1994) with their larger sample find that 14 of 18 RCB stars have quite 
similar compositions.  In this group, only hydrogen and lithium abundances vary strongly 
from star to star.
Nitrogen and sodium are also over-abundant.
Among the four RCB stars that have unusual compositions, V854 Centauri, V Coronae 
Australis, VZ Sagittarii, and V3795 
Sagittarii,  two are relatively hydrogen rich and all are iron 
poor like the LMC RCB stars (Lambert \& Rao 1994).
Glass et al. (1994) find long-term variations in the near-IR brightness in two of 
the LMC RCB stars and found a possible correlation between the IR brightness and decline 
activity for W Men.  
This behavior is similar to that seen in Galactic RCB stars.
Most also show an excess at IRAS wavelengths.  One LMC RCB star, HV 12842, seems 
to have been detected in the IRAS Faint Source Survey (Moshir et al. 1992) at a level of about 0.09 Jy at 12 
\micron. 
Despite some small differences in abundances and colors, 
the LMC RCB stars seem to be quite similar to their Galactic counterparts.

\section{New LMC RCB Stars}
\subsection{MACHO Photometry}
The MACHO Project (Alcock et al. 1992) is an astronomical survey experiment designed to 
obtain multi-epoch, two-color CCD photometry of millions of stars in the LMC (also, the 
Galactic bulge and SMC).  The survey makes use of a dedicated 1.27m telescope at Mount 
Stromlo, Australia and because of its southerly latitude is able to obtain observations of the 
LMC year round (Hart et al. 1996). The camera built specifically for this project (Stubbs et 
al. 1993) has a 
field of view of 0.5 square degrees which is achieved by imaging at prime focus. 
Observations are obtained in two bandpasses simultaneously, using a dichroic beamsplitter 
to direct the ``blue" ($\sim$4400-5900 \AA) and ``red" ($\sim$5900-7800 \AA) light onto 
2x2 mosaics of 2048x2048 Loral CCD's. Hereafter, these bandpasses will be referred to as 
$V_{MACHO}$ and $R_{MACHO}$, respectively. Images are obtained and read out 
simultaneously. The 15 \micron~pixel size maps to $0\farcs63$~on the sky. The data were 
reduced using a profile-fitting photometry routine known as SODOPHOT, derived from 
DoPHOT (Mateo \& Schechter 1989). This implementation employs a single starlist 
generated from frames obtained in good seeing.
The results reported in this survey comprise only a fraction of the planned data acquisition 
of the MACHO project. At present, most of the first three years' LMC data has been 
processed, consisting of some 5500 frames distributed over 22 fields; this sample contains 
a total of approximately 8 million stars. These data have been searched for variable stars 
and microlensing candidates and over 40,000 variables have been found, most newly 
discovered. The great majority of these fall into four well known classes: there are 
approximately 25,000 very red semiregular or irregular variables, 1500 Cepheids, 8000 
RR 
Lyraes, and 1200 eclipsing binaries (Cook et al. 1995). Typically, the dataset for a given 
star 
covers a timespan of about 1200 days and contains $\sim$700 photometric measurements 
(multiple observations are obtained on a given night whenever conditions allow).  The 
output photometry contains flags indicating suspicion of errors due to crowding, seeing, 
array defects, and
radiation events. 

The database of MACHO variables was searched for stars which underwent large sudden 
brightness variations.  The lightcurves of these large amplitude variables were then viewed 
by eye. 
Candidates were selected as having distinctive RCB lightcurve behavior.
RCB stars are true irregular variables (Clayton, Whitney \& Mattei 1993).
A star may have several declines in one year or go ten years or more without any declines.  
So any search over a short time period will detect only a fraction of the RCB stars.  
One year of MACHO photometry has been searched so far. 
Two candidates, MACHO*05:33:49.1-70:13:22 and  MACHO*05:32:13.3-69:55:59, have 
been 
found which show lightcurves characteristic of RCB stars.  
These coordinates are J2000.
Finding charts for the two stars are shown in Figures 1 and 2.
The fields are 161\arcsec~square, north is up and east to left.
The lightcurves are shown in Figures 3 and 4.  These figures include all available MACHO 
data up to the present.
Only data free from suspected errors are plotted, resulting in output photometry lists of 
length 394 (MACHO*05:33:49.1-70:13:22)  and 462 (MACHO*05:32:13.3-69:55:59). 
Typical photometric uncertainties are in the range 1.5-2~\%.
The $V_{MACHO}$ and $R_{MACHO}$ bandpasses have been converted to 
Kron-Cousins (KC) V and R bandpasses using the latest transformations determined from 
the ongoing internal calibrations of the MACHO database.
The $(V-R)_{KC}$ colors are also plotted in Figures 3 and 4.
The Color-Magnitude Diagrams for the fields of the two stars are shown in Figures 5 and 
6.  Both stars lie among the post-AGB stars.

\subsection{Spectroscopic Data}
Spectroscopic observations were obtained of MACHO*05:33:49.1-70:13:22 and  
MACHO*05:32:13.3-69:55:59 in 1995 November when both stars were at 
light maximum.  
The spectra were obtained with the Reticon photon-counting system
on the image-tube spectrograph on the SAAO 1.9m telescope at Sutherland, South Africa.
The grating used gives a reciprocal dispersion of 100 $\AA~mm^{-1}$ and a
resolution of approximately 4 \AA, giving a useful range of about
3600-5200 \AA~at the angle setting used.  The spectrograph is a two-aperture
instrument recording the star and sky simultaneously.  Normal operating procedure is to 
measure
the star through one aperture and then the other, so the sequence
goes arc, star in A, arc, star in B, arc. Each star is then wavelength
calibrated by the two arcs on either side and the results of star in
A and B are added together after flat-field correction and sky subtraction.
Flux calibration is done by observing one
standard star each night.  In the case
of MACHO*05:33:49.1-70:13:22, all the spectra were added together and a flux standard
from one night was used. The flux calibration is not accurate because
the instrument is not a spectrophotometer and observations are sometimes made in 
non-photometric conditions so that light losses vary with time and seeing. 
Although the absolute calibration is uncertain,
the relative fluxes should be reliable. MACHO*05:32:13.3-69:55:59 was observed on one 
night (2x1500 s) and MACHO*05:33:49.1-70:13:22 on three nights (2x1000 s, 2x1200 s, 
2x1000 s).
The spectra are shown in Figures 7 and 8.
These spectra are sums of all individual scans.

\section{Discussion}

Figure 3 shows the lightcurve of MACHO*05:33:49.1-70:13:22.  Almost the entire 
1200~d 
of coverage involves one deep decline of $\Delta V \ge$ 4 mag.  
The decline begins around JD 2448925 with a steep drop of $\sim$4 mag in a few days.  
There is a slight recovery around JD 249000 followed by another fading and then a slow 
recovery to maximum light.
This lightcurve is typical of an RCB star decline.  The final recovery to maximum light can 
be gradual as the dust cloud disperses and sometimes takes several years (e.g. Alexander et 
al. 1972).  
Figure 4 shows that lightcurve of  MACHO*05:32:13.3-69:55:59.  It is quite active 
showing 3 major declines around JD 2448900, 2449325 and 2449650.  There is a great 
variation in decline activity from star to star and also from time to time for an individual star 
(Clayton 1996).  The Galactic RCB star, V854 Cen, has shown similar activity to 
MACHO*05:32:13.3-69:55:59 in the last few years (Lawson et al. 1992).  
Both MACHO*05:33:49.1-70:13:22 and MACHO*05:32:13.3-69:55:59 show small 
amplitude variations at maximum light similar to other RCB stars.

Figures 3 and 4 also show the $(V-R)_{KC}$ color behavior.  
MACHO*05:33:49.1-70:13:22 becomes redder at the beginning of the decline and returns 
to its normal color as it 
returns to maximum light.  Early and late in a decline the star is reddened by a dust cloud 
which is not optically thick.  Deep in a decline the cloud may be optically thick so 
reddening may not be seen.
MACHO*05:32:13.3-69:55:59 shows $(V-R)_{KC}$ colors which become bluer at the 
onset of the decline and then return to normal as the star recovers to maximum.  RCB stars 
experience red and blue declines (Cottrell, Lawson, \& Buchhorn 1990).  The colors can 
vary from decline to 
decline depending on how much of the photosphere and the emission-line 
regions are obscured by dust, by the optical depth of the dust, and by the
relative strength of the emission lines. 
Sometimes very early in a decline, the colors are unchanged at first and then become bluer.  
This can occur if the forming cloud is smaller than the
photosphere and some unreddened starlight is still visible (Cottrell et al. 1990).  
Red declines occur if the forming cloud covers the entire photosphere. 
It is notable that MACHO*05:32:13.3-69:55:59 has had three blue declines in a row.  
Unfortunately, the data for Galactic RCB stars are sparse so the relative frequency of red 
and blue declines is not known.
For the LMC RCB stars, another possibility is confusion with a blue star in the aperture although no such star is visible on the CCD frames.

The spectrum of MACHO*05:33:49.1-70:13:22, shown in Figure 7, is very similar to that 
of the hot RCB star, V348 Sgr (Dahari \& Osterbrock 1984; Leuenhagen \& Hamann 
1994).  
The spectrum of V348 Sgr is classed as WC11 since it shows emission at C II but not C III 
(Leuenhagen \& Hamann 1994).
In addition, its lightcurve, IR excess and hydrogen deficiency distinguish it as an RCB 
star.
The  MACHO*05:33:49.1-70:13:22 spectrum shows strong C II emission at 3919, 4267 
and 4735 to 4747 \AA.  
There is also possible C II emission seen at 4618 to 4630 \AA~and near 4861 \AA~blended 
with $H\beta$.  
In addition, MACHO*05:33:49.1-70:13:22 seems to have been detected in the IRAS 
Serendipitous Survey (Kleinmann et al. 1986) at a level of 0.1 Jy at 12 \micron.
This is similar to the flux detected for HV 12842 and both are consistent with an 
extrapolation of flux levels measured for Galactic RCB stars.
The spectrum, lightcurve and IR excess of 
MACHO*05:33:49.1-70:13:22 indicate that it is a hot RCB star, only the fourth known and 
the first discovered outside the Galaxy.  In addition, although the spectrum is low 
resolution the C II lines show a redshift of 259$\pm$31 $km~s^{-1}$ which is appropriate for LMC membership.  

The spectrum of MACHO*05:32:13.3-69:55:59 is shown in Figure 8.  This spectrum is a 
stereotypical cool RCB spectrum similar to S Aps and V517 Ophiuchi (Kilkenny et al. 
1992).
The spectrum shows deep Swan bands of $C_2$ with bandheads at 4382, 4737, and 5165 
\AA, and violet bands of CN with bandheads at 3883, 4216 and 4606 \AA.  In addition, 
the spectrum is distinguished as an RCB star by the weak or absent Balmer 
lines, G-band and $^{12}C^{13}C$ bands (Lloyd Evans, Kilkenny, \& van Wyk 1991).  
This is a result of severe hydrogen deficiency and a lack of $^{13}C$ typically seen in 
RCB stars (Clayton 1996).  MACHO*05:32:13.3-69:55:59 shows a $C_2$ (1,0) 4737 
\AA~band that dips about 80\% 
below the continuum. 
HV 5637 shows a 72\% depression and S Aps about 79\% (Feast 1972). 
W Men and HV 12842 show much smaller dips of about 10\% much like R CrB and RY 
Sgr.  
The (V-R$)_{KC}$ colors of S Aps and MACHO*05:32:13.3-69:55:59 are similar.
The spectrum, color, and lightcurve of MACHO*05:32:13.3-69:55:59 show that it is a cool 
RCB 
star.
The measured positions of the molecular bandheads show a redshift of 269$\pm$21 $km~s^{-1}$ which is appropriate for
LMC membership.

As mentioned in the introduction, the distance scale for the RCB stars depends entirely on 
the LMC members.  Using the recent photometry of Lawson et al. (1990), we find 
$V_{max}$ = 14.8, 13.8, and 13.7 for HV 5637, W Men and HV 12842, respectively.
On the basis of this small sample, RCB stars are supergiants with a range of $\sim$1 
magnitude in absolute luminosity in the V-band.  Since HV 5637 is cooler than W Men and 
HV 12842, Feast (1979) suggested a relationship between temperature and luminosity.  
The observed range in $V_{max}$ is likely to be intrinsic rather than due to reddening 
differences.
Estimates of foreground (circumstellar and interstellar) reddening are somewhat uncertain 
since RCB star colors are not known a priori.
Older studies of the Galactic foreground find a fairly uniform screen of dust of E(B-V) 
$\sim0.04 - 0.07$ (e.g. McNamara \& Feltz 1980).   
Schwering \& Israel (1991) re-examined the foreground reddening by comparing H I and 
IR observations. They find a small but significant variation in foreground reddening across 
the face of the LMC from E(B-V) = 0.07 to 0.17 mag.
Any constant component of circumstellar dust around RCB stars is small.  
Using the observed B-V and the estimated $T_{eff}$ listed in Table 2, we calculate E(B-V) 
$\sim$ 0.1-0.2 for HV 5637, W Men and HV 12842 (Johnson 1966). 
Goldsmith et al. (1990) estimate 0.08 and 0.10 for the circumstellar and interstellar 
components of the E(B-V) towards W Men.  Only W Men has a measured $(V-R)_{KC}$ 
(Eggen 1970).
Converting this to Johnson $(V-R)_J$, a similar E(B-V) is obtained assuming a 
normal extinction curve with $R_v$=3.1 (Cousins 1980).

Therefore, the new stars presented here are very important.  The measured $V_{max}$ for 
these stars are 16.1 and 16.3, fainter by about 2 mag than the three known RCB stars.  Do these values really represent unreddened $V_{max}$?  
These stars have been followed photometrically for only about 3 years so it is possible that 
they have never fully recovered to maximum light.
RCB stars go through very active phases where they are in decline for years (e.g. Mattei, 
Waagen \& Foster 1991).  However, to remain in decline, dust must form regularly to 
compensate for the dispersal of previous dust clouds.
Flat decline-lightcurve behavior only occurs deep in a decline when all direct starlight is 
extinguished and only scattered light is seen.  This kind of behavior is not seen when 
$\Delta V \sim$ 2 mag so it can't account for the observed lightcurve.
Therefore, the recent lightcurve behavior of both stars, seen in Figures 3 and 4, is 
consistent with being at or near maximum light.  MACHO*05:33:49.1-70:13:22 seems to 
be approaching maximum light after a long decline.  This lightcurve shape is seen in many 
other RCB declines.  The MACHO*05:32:13.3-69:55:59 lightcurve is more complicated 
and the star is definitely in an active phase.  However, in the last 200 d during which time 
the spectrum was obtained, the star has been very constant with $\Delta V \sim$ 0.25 mag.  
There is no evidence in this time period for either dust formation or dispersal.

Another possibility is large interstellar extinction toward these stars.  The reddening can be 
estimated from the measured $(V-R)_{KC}$ colors which were converted to $(V-R)_J$ 
(Cousins 1980).  Assuming normal supergiant $(V-R)_J$ colors then E(B-V) $\sim$0.3 
for MACHO*05:33:49.1-70:13:22 and  $\sim$0.4 for MACHO*05:32:13.3-69:55:59.  
This is slightly higher than the estimate of $\sim$0.1-0.2 for the other LMC RCB stars. 
These values of E(B-V) are seen for many other LMC stars although most lie in or near the 
30 Dor region.  Neither of the new LMC RCB stars lies in a visibly dusty region of the 
LMC.
Estimates of $V_o$ and $M_V$ for all five stars are given in Table 2 using the estimated 
values of E(B-V).  The new stars fall $\gtrsim$ 0.5 mag below the range of $M_V$ = -4.1 
to -5.5 for the previously known LMC RCB stars.
These estimates are somewhat uncertain because RCB colors may not be the same as 
normal supergiants.  
For instance, S Aps has the same maximum-light $(V-R)_{KC}$ color as 
MACHO*05:32:13.3-69:55:59 and a similar $T_{eff}$ (Lawson et al. 1990) yet its B-V 
colors imply a smaller reddening of E(B-V) $\sim$0.1.  If the reddening of 
MACHO*05:32:13.3-69:55:59 is 0.1 then it has an even fainter $M_V \sim$-2.6.  
Another indication of the uncertain reddening correction can be seen in Figures 5 and 6.  
The stars in the MACHO*05:33:49.1-70:13:22 field seem to be $\sim$ 0.1 mag redder than 
the 
stars in the MACHO*05:32:13.3-69:55:59 field.
Therefore, taking into account the uncertainties in the intrinsic and measured colors, and 
stellar $T_{eff}$, the uncertainty in E(B-V) is $\sim$0.1-0.2.  So the uncertainty in 
$M_V$ is $\sim$0.3-0.6.

A slightly smaller range of values is found when using $M_{Bol}$. 
The assumed effective temperatures are 5000 and 7000 K for cool and warm RCB stars, 
respectively.  
Using the values of $M_V$ calculated above and Bolometric corrections for normal 
supergiants, we get the values listed in Table 2 for $M_{Bol}$.
A value of 20,000 K has been estimated for V348 Sgr (Sch\"{o}nberner 
1986).  This value has been applied to MACHO*05:33:49.1-70:13:22.    
It's $M_{Bol}$ lies in the range of the warm RCB stars.
The two cool RCB 
stars have the lowest Bolometric luminosities.
The question of whether LMC and Galactic RCB stars are intrinsically different 
remains open.  There are abundance differences but they may lie within the range of 
variations seen in the Galactic RCB stars.  The slightly bluer UBV colors of the LMC RCB 
stars may 
also be an indication of abundance differences.

\section{Summary}

$\bullet$ The discovery of new LMC RCB stars brings to five the number known.  Most 
importantly, one is a rare hot RCB star.

$\bullet$ The wider range of $T_{eff}$ now existing does not support the suggestion of 
a relationship between $T_{eff}$ and $M_V$.  However, the two cool RCB stars have the 
lowest Bolometric luminosities.

$\bullet$ The new stars presented here suggest that there is a wider range of absolute visible
magnitude than given by the canonical $M_V \sim$ -4 to -5. 
Therefore, the absolute luminosities of RCB stars are now less certain since they are based 
solely on the LMC stars.

$\bullet$ This pilot project shows the value of the MACHO photometric database for the 
discovery of new RCB stars in the LMC.  Based on the success of this project, more RCB 
stars will be found in the future.  Increasing the sample of stars is the only way to resolve 
the uncertainty about the absolute luminosities of the RCB stars.

\clearpage

\begin{deluxetable}{llll}
\small
\footnotesize
\tablewidth{0pt}
\scriptsize
\tablecaption{Luminosity Estimates for Galactic RCB stars }
\tablehead{\multicolumn{1}{c}{Name}  
&
\multicolumn{1}{c}{$M_{V}$}& 
\multicolumn{1}{c}{Method}&
 \multicolumn{1}{c}{Ref}}
\startdata
R CrB&-3.1&Member Wolf 630 Group&Eggen (1969)\nl
R CrB&-4.6&Mg II Emission Core&Rao et al. (1981)\nl
RY Sgr&-4&Close Companion&Andrews et al. (1967)\nl
SU Tau&-3&I.S. Polarization&Doroshenko et al. (1978)\nl
V482 Cyg&-2.8&Close Companion&Gaustad et al. (1988)\nl
V482 Cyg&-4.6&Radial Velocity&Rao \& Lambert (1993)\nl
V348 Sgr&-4.8&Radial Velocity&Sch\"{o}nberner (1986)\nl
\enddata
\end {deluxetable}

\begin{deluxetable}{lllllllll}
\small
\footnotesize
\tablewidth{0pt}
\scriptsize
\tablecaption{LMC RCB stars }
\tablehead{\multicolumn{1}{c}{Name}  
& \multicolumn{1}{c}{$V_{Max}$}&
 \multicolumn{1}{c}{$B-V$}&
 \multicolumn{1}{c}{$(V-R)_{KC}$}&
 \multicolumn{1}{c}{$E(B-V)$}&
 \multicolumn{1}{c}{$V_o$}&
\multicolumn{1}{c}{$M_{V}$}& 
\multicolumn{1}{c}{$M_{Bol}$}& 
\multicolumn{1}{c}{$T_{eff}$}}
\startdata
HV 5637&14.8&1.25&\nodata&0.1&14.5&-4.1&-4.6&5000\nl
W Men&13.8&0.37&0.23&0.1&13.5&-5.1&-5.3&7000\nl
HV 12842&13.7&0.50&\nodata&0.2&13.1&-5.5&-5.7&7000\nl
MACHO*05:33:49.1-70:13:22&16.1&\nodata&0.1&0.3&15.2&-3.4&
-5.4&20,000\nl
MACHO*05:32:13.3-69:55:59&16.3&\nodata&0.8&0.4&15.1&-3.5&-4.0&5000\nl
\enddata
\end {deluxetable}

\clearpage
\begin{figure}
\plotone{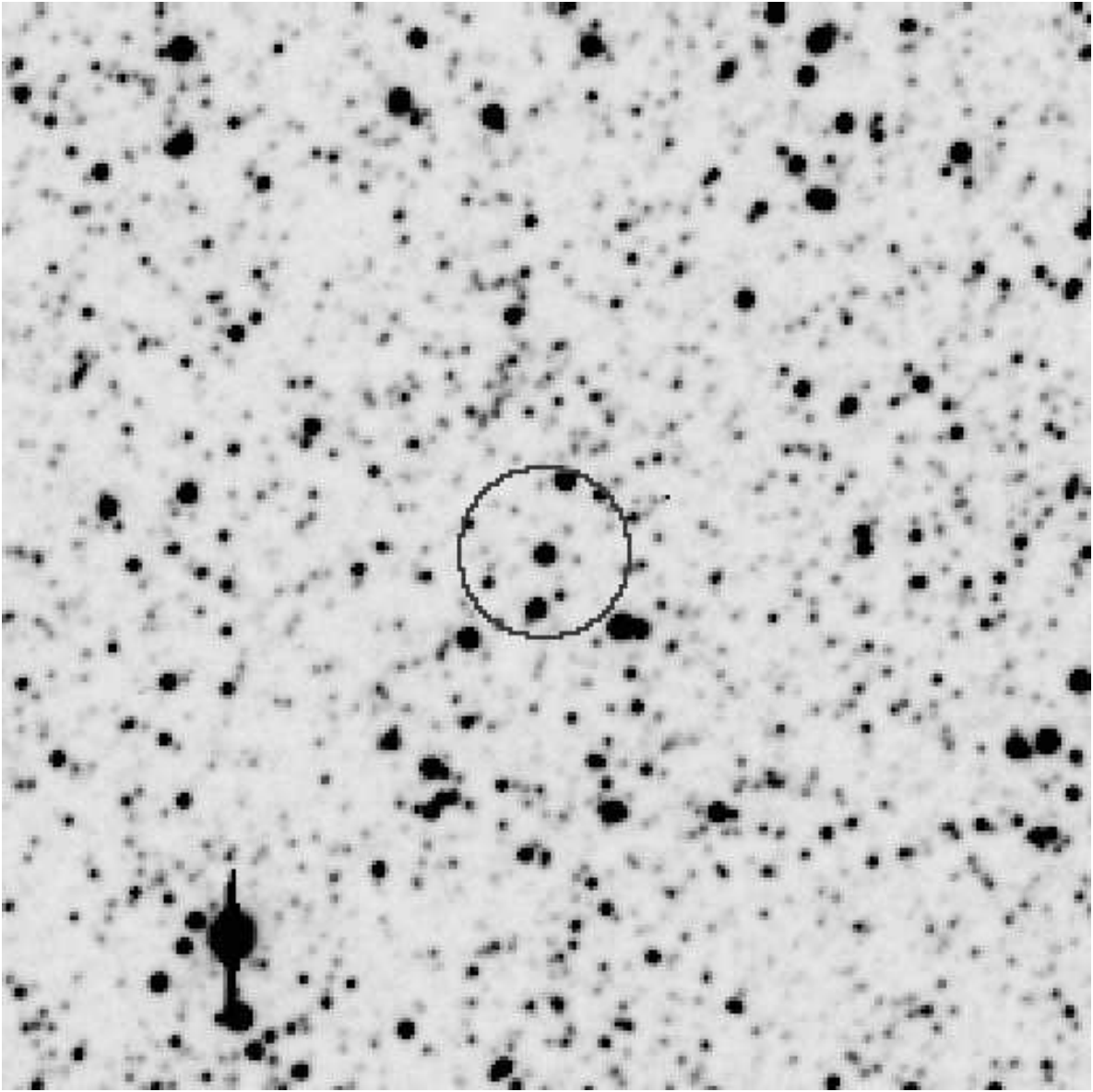}
\caption{Finding chart for MACHO*05:33:49.1-70:13:22.
The field is 161\arcsec~square, north is up and east to left.}
\end{figure}

\clearpage

\begin{figure}
\plotone{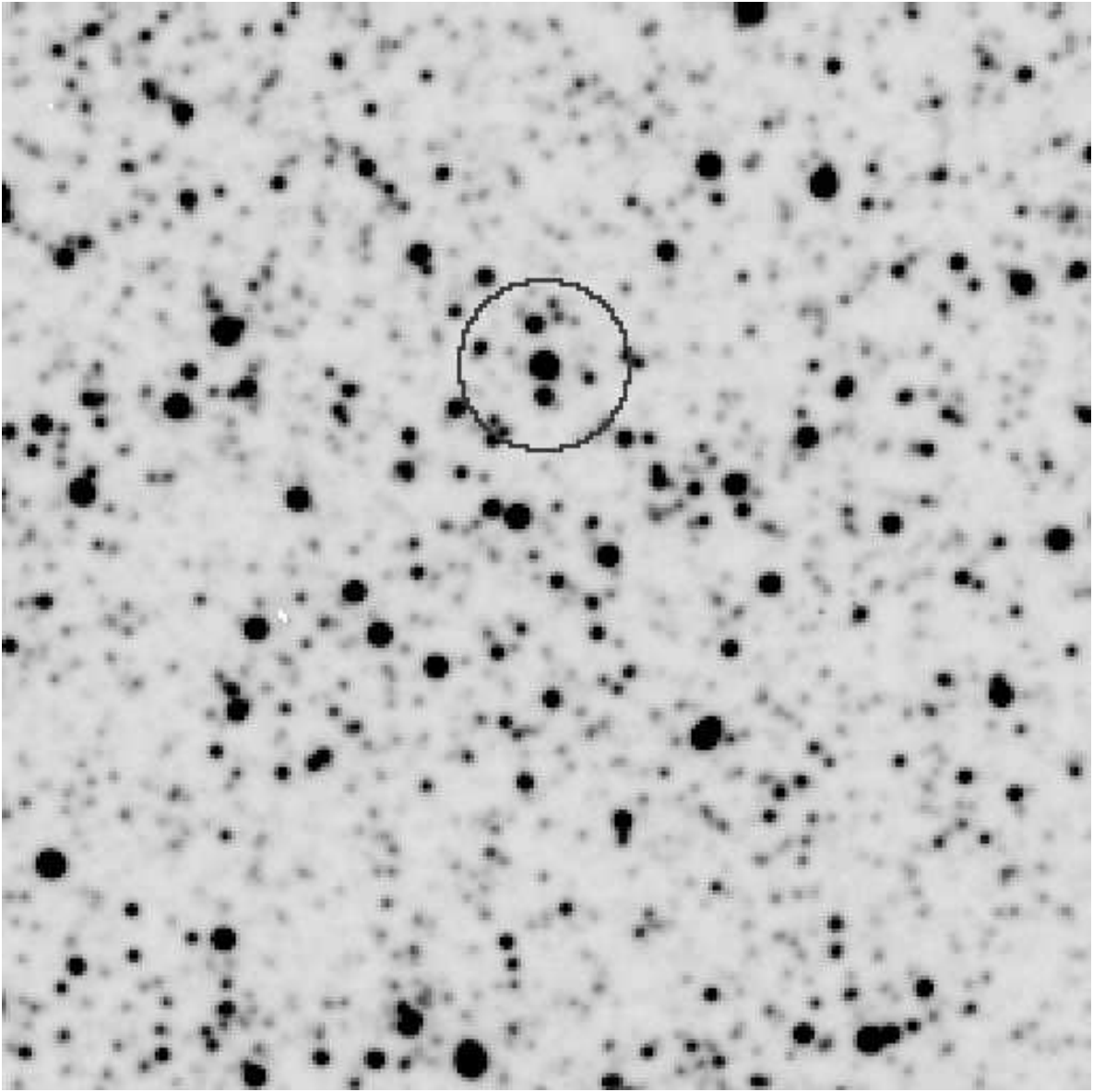}
\caption{Finding chart for MACHO*05:32:13.3-69:55:59.
The field is 161\arcsec~square, north is up and east to left.}
\end{figure}

\clearpage

\begin{figure}
\plotone{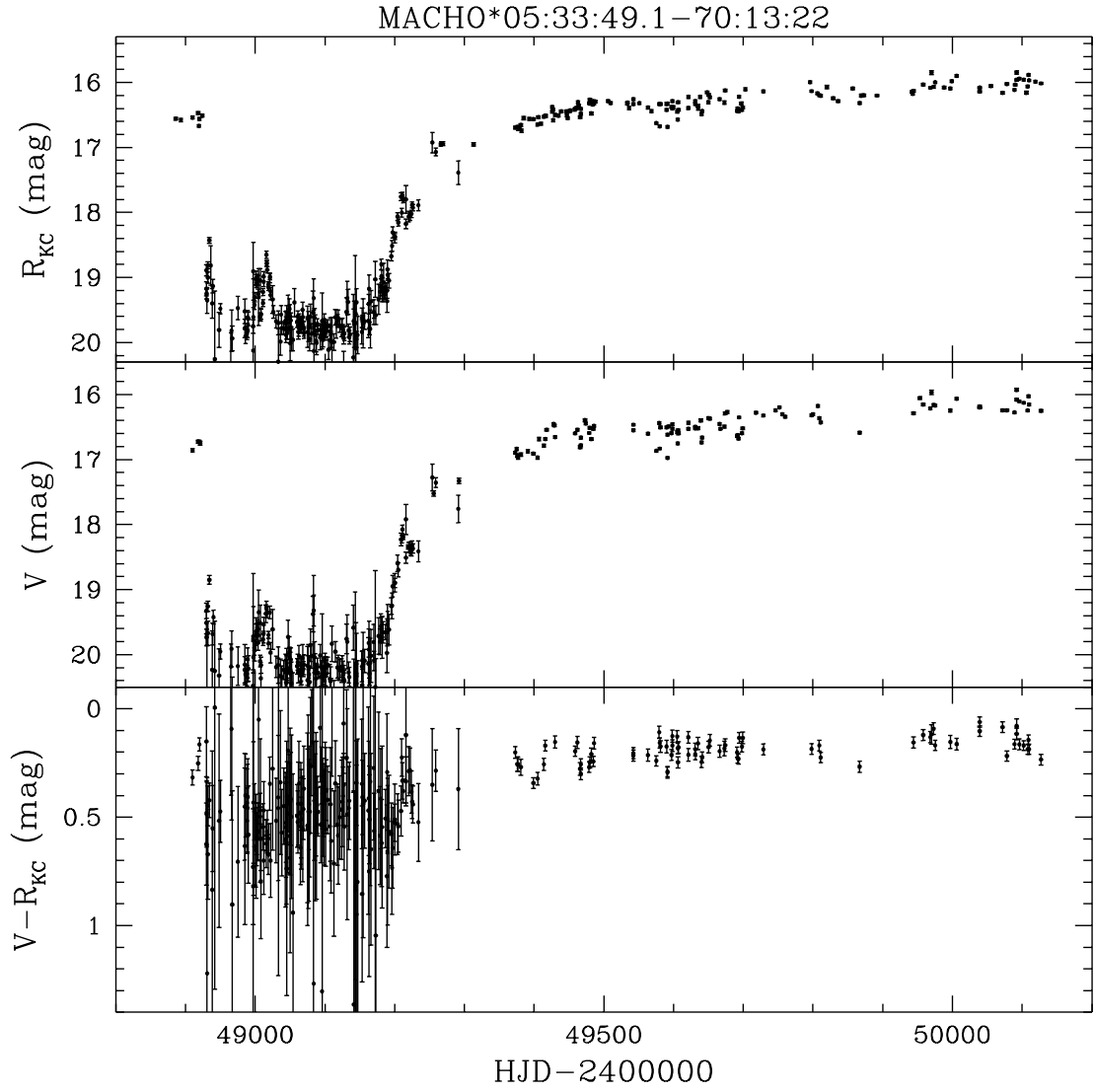}
\caption{MACHO photometry for MACHO*05:33:49.1-70:13:22.  
The data have been converted to Kron-Cousins V- and R-bands.  
See text. }
\end{figure}

\clearpage

\begin{figure}
\plotone{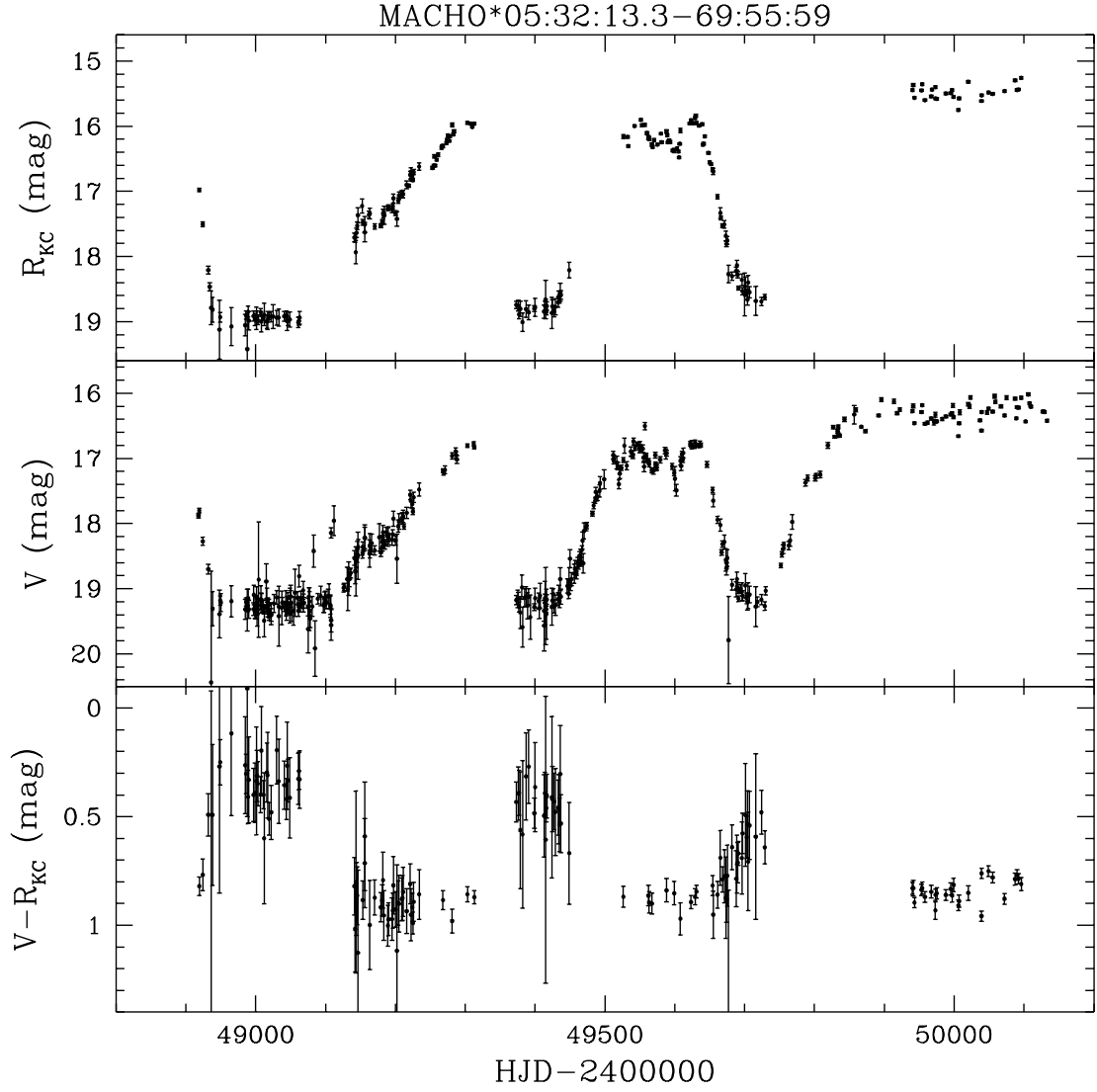}
\caption{MACHO photometry for MACHO*05:32:13.3-69:55:59. The
data have been converted to Kron-Cousins  V- and R-bands.  
See text. }
\end{figure}

\clearpage

\begin{figure}
\plotone{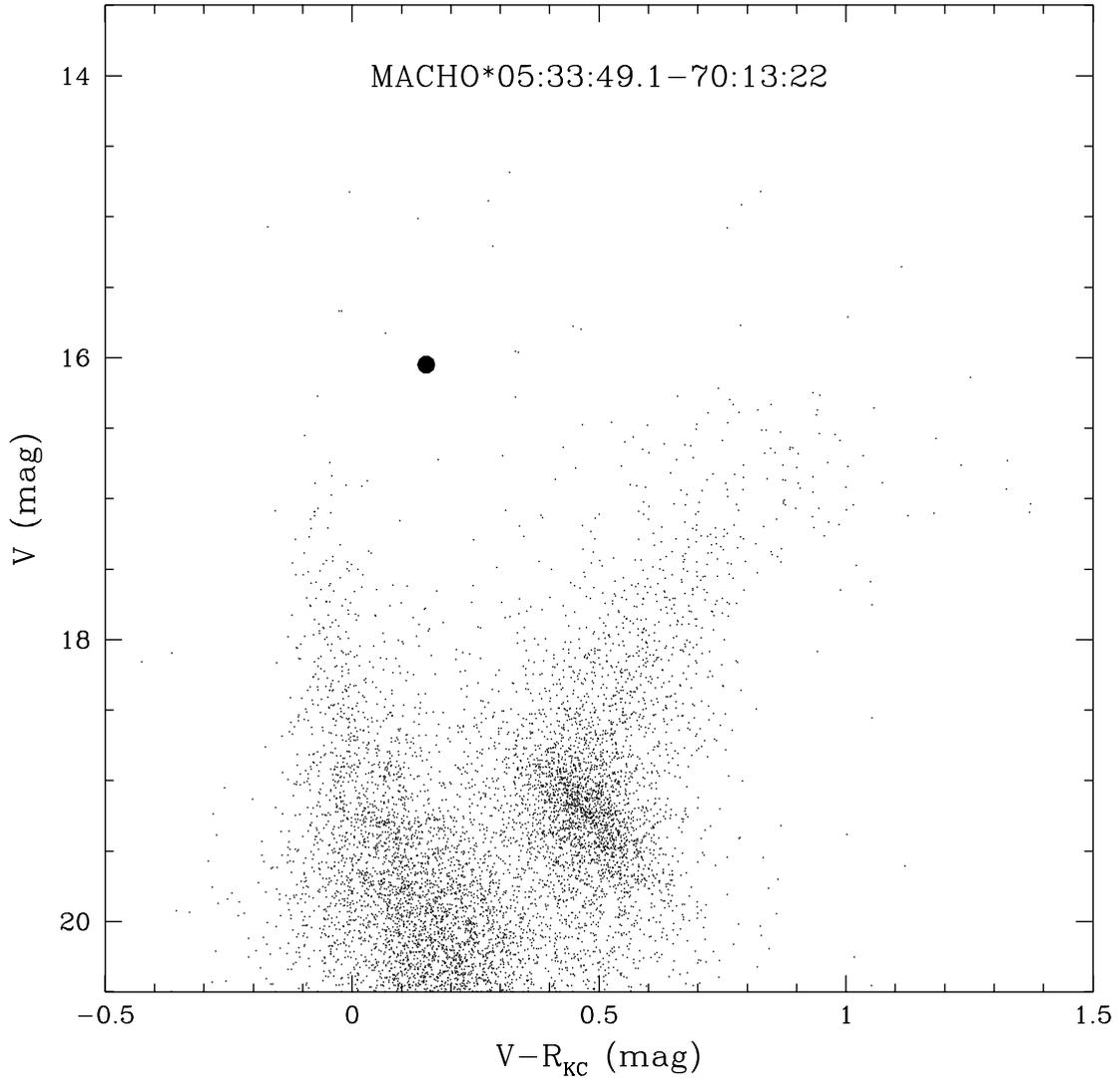}
\caption{Color-Magnitude Diagram for stars in the
MACHO*05:33:49.1-70:13:22 field.  The star itself is plotted 
as the large filled circle.}
\end{figure}

\clearpage

\begin{figure}
\plotone{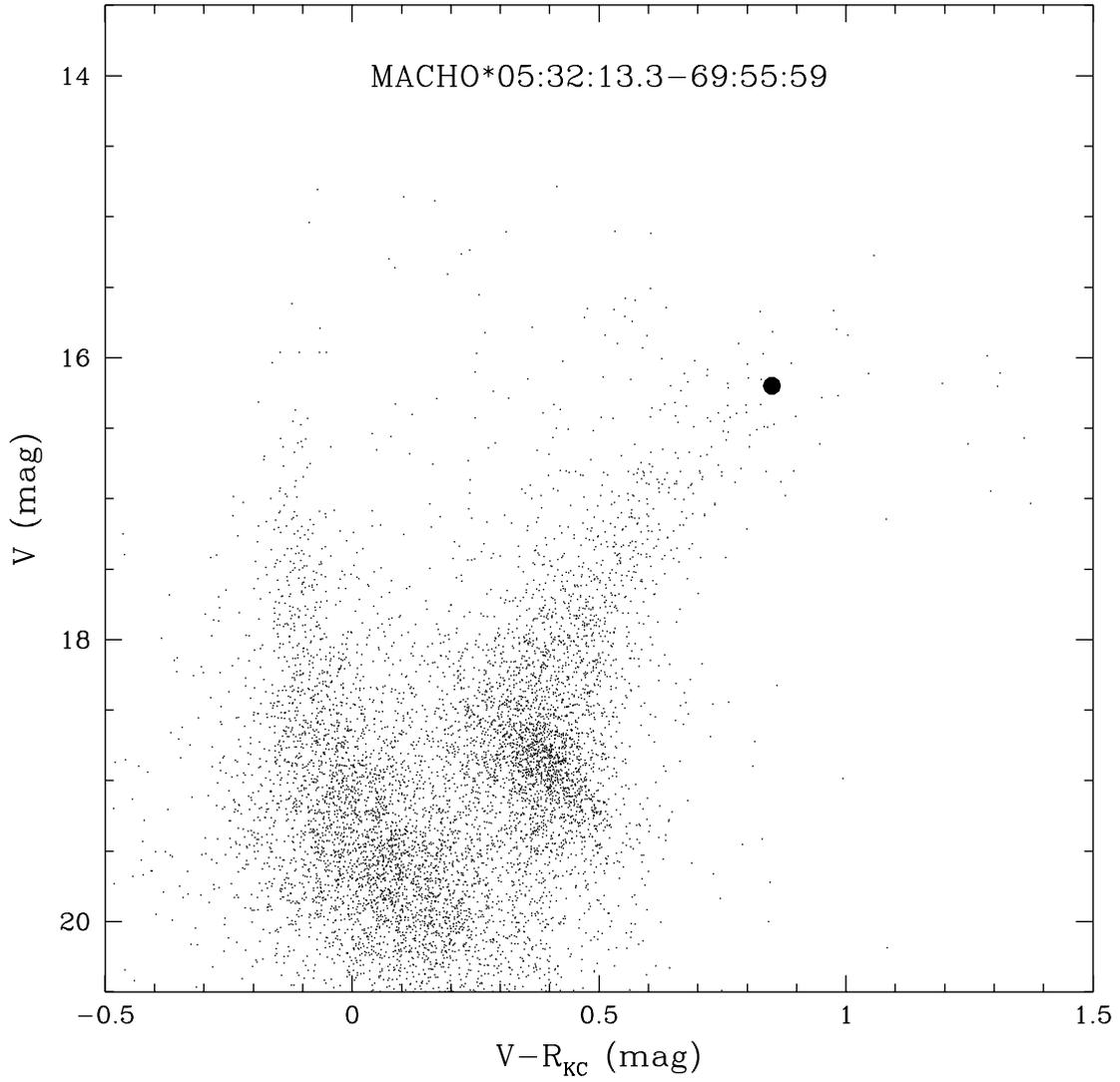}
\caption{Color-Magnitude Diagram for stars in the 
MACHO*05:32:13.3-69:55:59 field. The star itself is plotted 
as the large filled circle.}
\end{figure}

\clearpage

\begin{figure}
\plotone{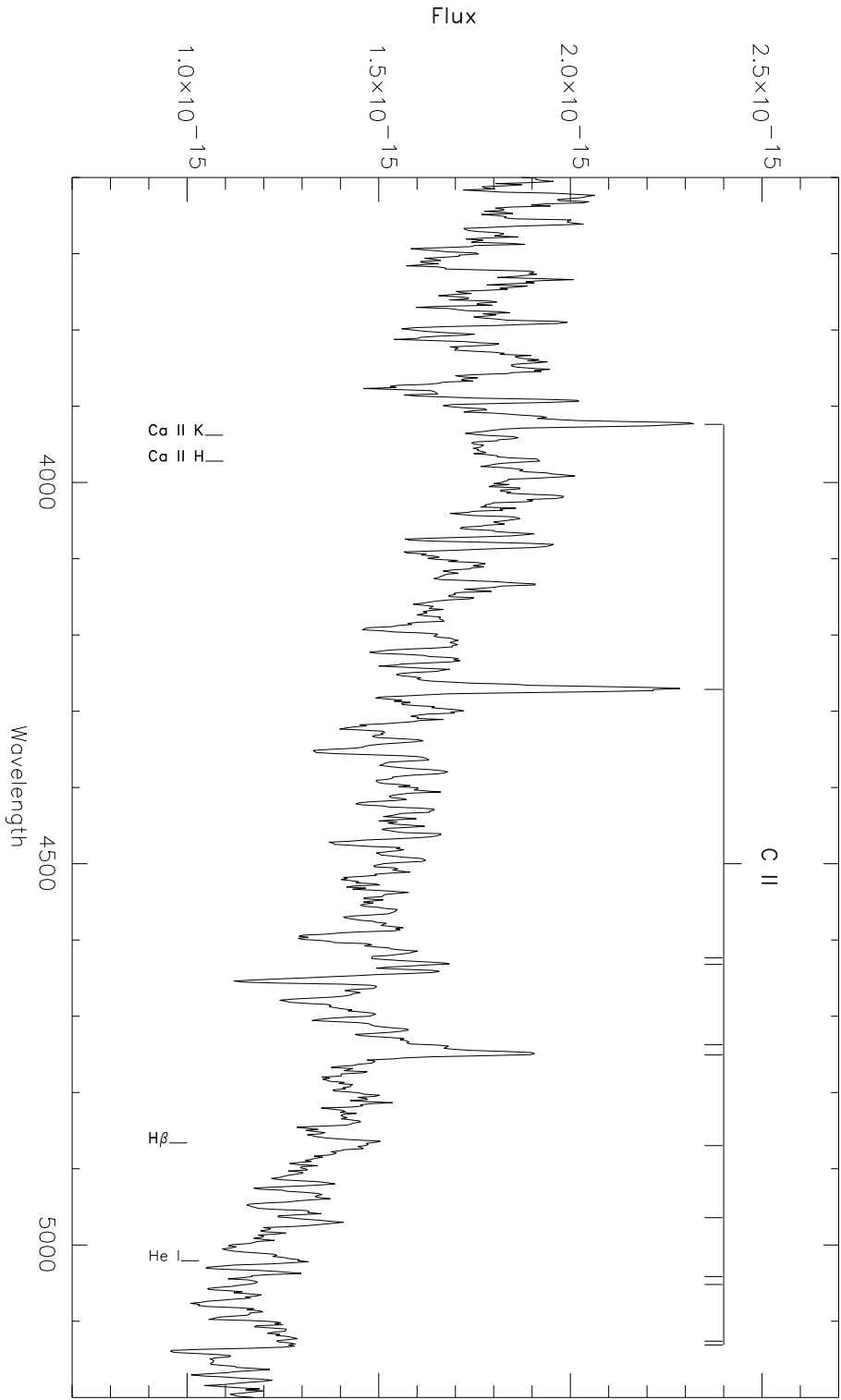}
\caption{Maximum light spectrum of MACHO*05:33:49.1-70:13:22.}
\end{figure}

\clearpage
 
\begin{figure}
\plotone{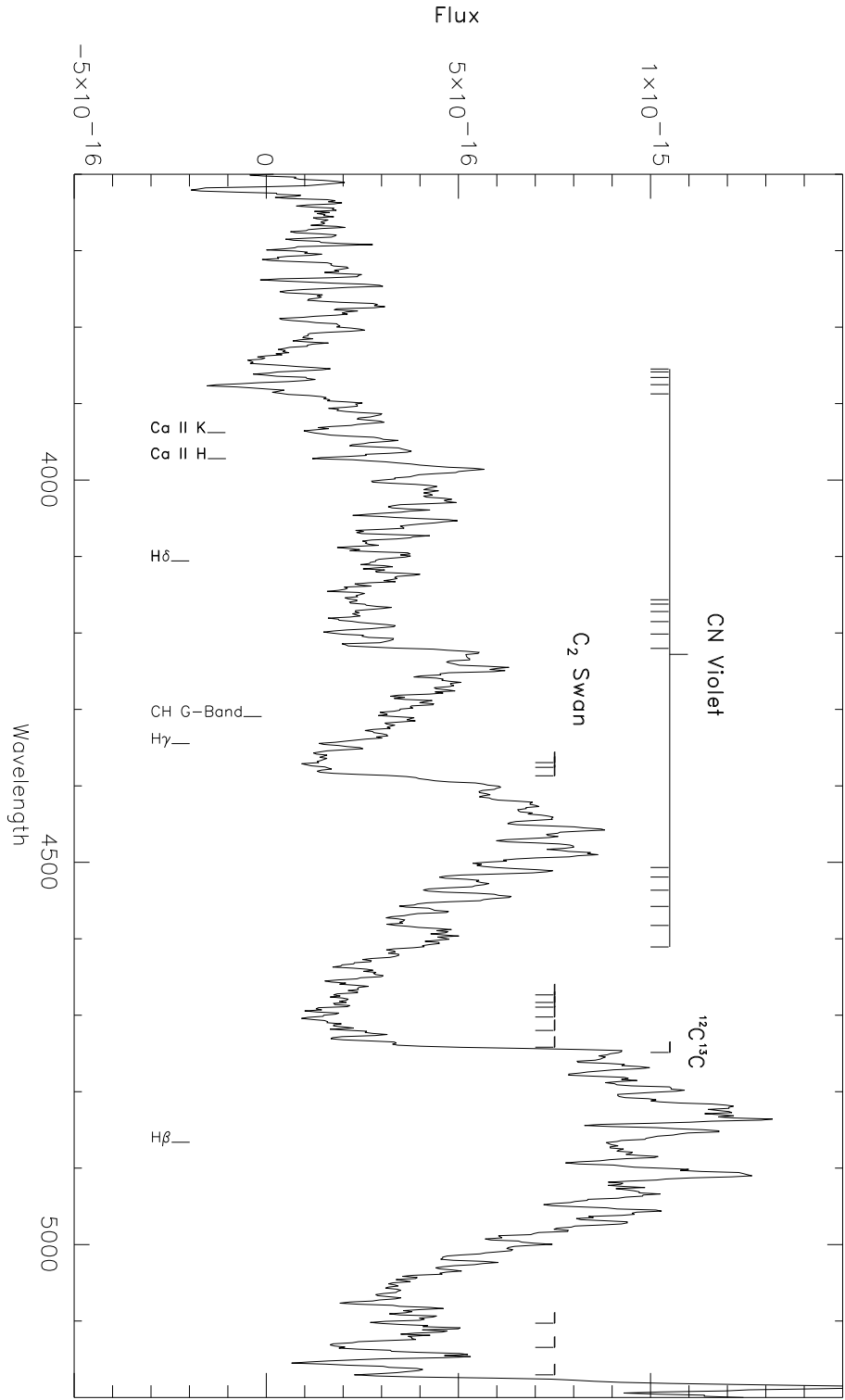}
\caption{Maximum light spectrum of MACHO*05:32:13.3-69:55:59.}
\end{figure}

\end{document}